\documentclass[preprint]{aastex}
\begin{document}
\title{Is the X-ray variable source in M82 due to gravitational lensing ?}
\author{Da-Ming Chen}
\affil{National Astronomical Observatories, Chinese Academy of
Sciences, Beijing 100012, China}

%\maketitle

\begin{abstract} 
We explore the possibility of attributing the recent discovery of 
the hard X-ray variable source CXOM82 J095550.2+694047
in M82 to the gravitational magnification by 
an intervening stellar obeject along the line of sight as microlens. 
The duration of the event ($>84$ days) allows us to set robust
constraints on the mass and location of the mircolensing object
when combined with the dynamical properties of the Galactic halo,
the M82 and typical globular clusters.  Except for the extremely low
probablity, the microlensing magnification by MACHO in either the 
Galactic halo or the M82 halo is able to explain the X-ray variability of
CXOM82 J095550.2+694047. It is hoped that the lensing
hypothesis can be soon tested by the light curve measurement.
\end{abstract}

\section{Introduction}

Observations of the most famous starburst galaxy
M82 with the High-Resolution Camera on board the {\it Chandra
X-Ray Observatory}, show that there are nine sources in the
central $1'\times 1'$ region, but no source was detected at the
galactic center\citep{mat01}. Comparing the observations on 1999
October 28 and on 2000 January 20, the authors found an extremely
large time variability of the source CXOM82 J095550.2+694047,
which is $9''$ away from the galactic center. They concluded that
this source is the origin of the hard X-ray time variability of
M82 detected with ASCA. Assuming a spectral shape obtained by the
ASCA observation, its luminosity in the 0.5 - 10 KeV band changed
from $1.2\times 10^{40}$ erg s$^{-1}$ on 1999 October 28 to
$8.7\times 10^{40}$ erg s$^{-1}$ on 2000 January 20.

It is difficult to explain such a short-term variability in terms of
a supernova remnant. If the spectral shape of the source is described
by an absorbed thermal bremstrahlung model, the observations show
that the probability of such a bright source existing in the
$1'\times 1'$ field is $~ 0.3\%$. Consequently, this source is 
probably not a background AGN.

CXOM82 J095550.2+694047 may be a medium-massive black hole
\citep{mat01}, and the possibilities that an X-ray binary source
whose jet is strongly beamed towards us cannot be excluded. In this
letter, we explore an alternative possibility that the time
variability of the source is originated from the microlening effect
by MACHO along the line of sight.

\section{Microlensing explanation}

The source CXOM82 J095550.2+694047 is $9''$ away from the center
of M82 (its distance from us is 3.9 Mpc). It may be located in a
star cluster with typical size of 2.34 - 10 pc \citep{de01}and  typical
velocity dispersion of member stars of 15 km/s.

If a foreground star is traversing across the line of sight, the
apparent magnitude of the source would be magnified according to
microlensing theory. The duration of the crossing time is
$T=2a_E/v$, in which $a_E$ is the Einstein radius defined as
\begin{equation}
a_E=\left(\frac{4GM}{c^2}\frac{D_{ds}D_d}{D_s}\right)^{1/2},
\end{equation}
where $D_d$, $D_s$ and $D_{ds}$ are the angular diameter distances
to the lens, to the source and from the lens to the source,
respectively, $v$ is the relative velocity perpendicular to the
line of sight between the source and the lens.

The mass of microlens is related to the crossing time as \citep{wu96}
\begin{equation}
M=\frac{c^2}{32G}\frac{D_s}{D_dD_{ds}}\frac{v^2T^2}{0.4919},
\label{m2t}
\end{equation}
where we have chosen the total maximum magnification to be $\mu_{T}=7.0$,
since the luminosity of the source in the 0.5 - 10 KeV band changed
from $1.2\times 10^{40}$ erg s$^{-1}$ on 1999 October 28 to
$8.7\times 10^{40}$ erg s$^{-1}$ on 2000 January 20.

%\clearpage
A microlens of mass of $\sim M_{\odot}$ could lie
at any point between the source and the observer, and
different positions would lead to different crossing time. For the
source CXOM82 J095550.2+694047, the observed time duration must
exceed 84 days. For simplicity,  we can take $T=2\times 84$ days
for an approximate estimate. The microlens mass versus its distance
is plotted in Fig.1 for three kinds of lensing models: the microlens
is located in our Galaxy with a typical velocity of 200km/s, in M82 with
a typical velocity of 100km/s and in the star cluster of the source with
a velocity of 15km/s.

At three different positions of a microlens in the Galaxy, the M82
and the star cluster of the source,
the same microlens mass would lead to different crossing time durations,
as is shown in Fig.2.
We set $D_s=3.9$Mpc in all the panels. Because of the symmetry between $D_d$
and $D_{ds}$ in equation (\ref{m2t}), the first and second panels both in Fig.1
and Fig.2 look very similar.

\begin{figure}[r]
%\vspace{5cm}
\epsscale{}
\plotone{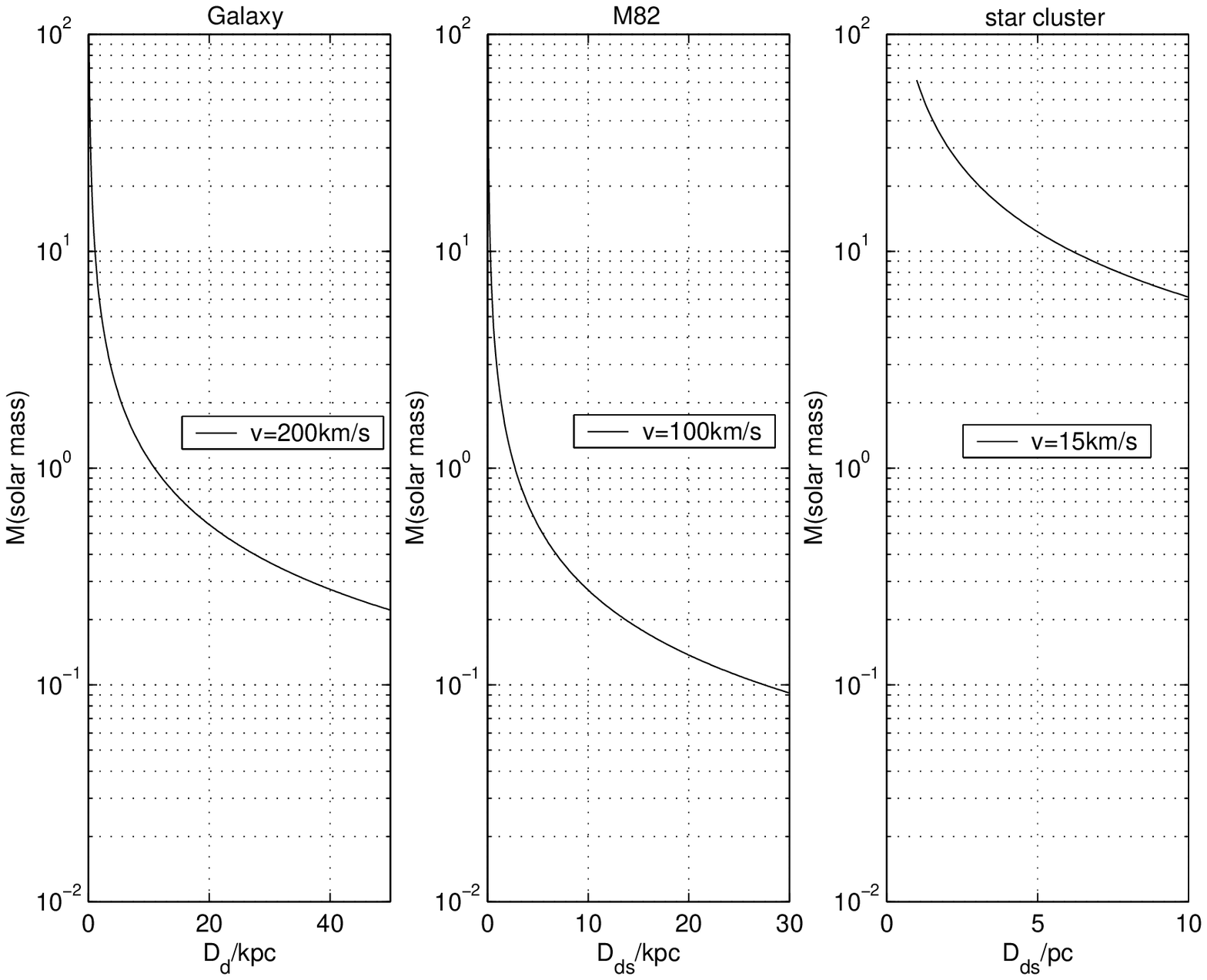}
\caption{The mass of microlens as a function of $D_d$ or $D_{ds}$. Three
panels from left to right refer to the positions of lenses located in our
Galaxy, in the M82 and in a star cluster, respectively. The corresponding
typical velocity of the microlens is clearly marked.}
\end{figure}

\begin{figure}[r]
\epsscale{}
\plotone{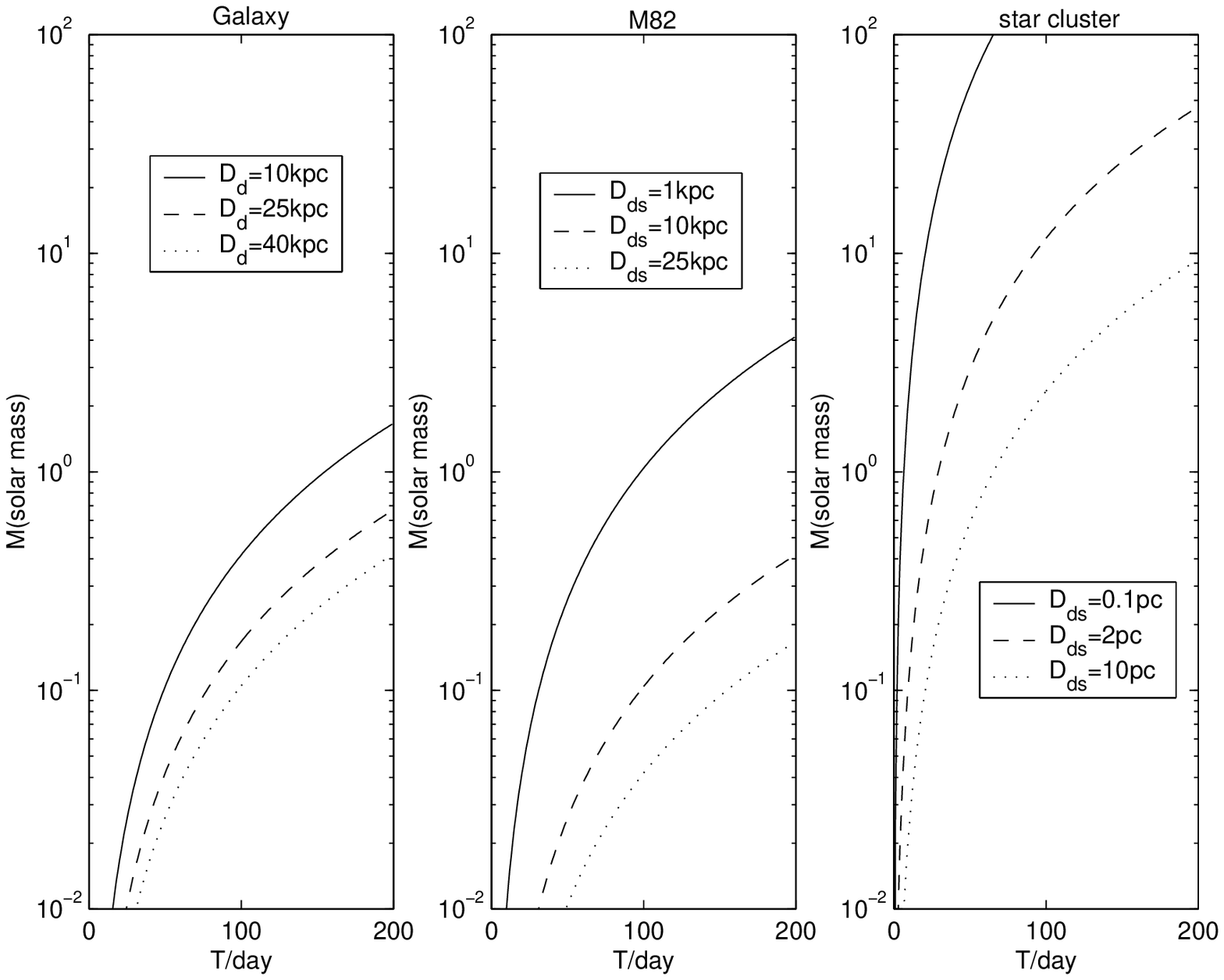}
\caption{The mass of microlens as a function of crossing time duration $T$.
Three panels from left to right refer to the positions of lenses located
in our Galaxy, in the M82 and in a star cluster, respectively.
The position of the microlens in each case is also indicated.}
\end{figure}

\section{Discussions and conclusions}

It can be seen clearly from Fig.1 and Fig.2 that the microlens object, if any,
should be located in M82 or the Galactic halo. However, we can exclude
the possibility of the self-lensing  by a star in the star cluster
where CXOM82 J095550.2+694047 may be harbored if the typical size of a
star cluster is smaller than  $\sim10$ pc.

The probability that a source is gravitationally lensed
is described by the optical depth which amounts to $\sim10^{-6}$.
Indeed, the event like  the hard X-ray variable source
CXOM82 J095550.2+694047 seen in M82 is very rare if it is due to
microlensing. A conclusive resolution needs a detailed sample of
the light curve of CXOM82 J095550.2+694047. Whether or not the
light curve demonstrates the time-symmetry around the maximum
intensity will be  a crucial test for the lensing hypothesis.

\acknowledgments

I'm grateful to Prof. Xiang-Ping Wu for reading through the manucript
and giving constructive suggestions. This work was supported by
the National Science Foundation of China, under grant 19725311
and 19873014.


\begin{thebibliography}{}
\bibitem[Matsumoto et al.(2001)]{mat01}Matsumoto, H. et al. 2001, \apjl,
547, 25
\bibitem[de Grijs et al.(2001)]{de01}de Grijs, R. et al. 2001, \aj,
121, 768
\bibitem[Wu(1996)]{wu96}Wu, Xiang-Ping 1996, Fundamentals of Cosmic Physics,
17,1
\end{thebibliography}
\end{document}